\documentclass[draft]{agujournal}

\journalname{JGR-Space Physics}

\begin{document}

%
%

\title{Explicit IMF $B_y$-dependence in high-latitude geomagnetic activity}

%
%

 \authors{L. Holappa\affil{1,2,3} and K. Mursula\affil{1}}

\affiliation{1}{ReSoLVE Centre of Excellence, Space Climate Research Unit, University of Oulu, Oulu, Finland.}
\affiliation{2}{Solar Physics Laboratory, NASA Goddard Space Flight Center, Greenbelt, MD, USA.}
\affiliation{3}{Department of Physics, The Catholic University of America, Washington, DC, USA.}


\correspondingauthor{L. Holappa}{lauri.holappa@oulu.fi}


\begin{keypoints}
\item IMF $B_y$ is an explicit driver of high-latitude geomagnetic activity.
\item High-latitude geomagnetic activity is suppressed in local winter for $B_y<0$ in Northern Hemisphere and for $B_y>0$ in Southern Hemisphere.
\item Explicit $B_y$-effect maximizes when the Earth's dipole axis points towards night.
\end{keypoints}

%
%

\begin{abstract}

The interaction of the solar wind with the Earth's magnetic field produces geomagnetic activity, which is critically dependent on the orientation of the interplanetary magnetic field (IMF). 
Most solar wind coupling functions quantify this dependence on the IMF orientation with the so-called IMF clock angle in a way, which is symmetric with respect to the sign of the $B_y$ component.
However, recent studies have suggested that the sign of $B_y$ is an additional, independent driver of high-latitude geomagnetic activity, leading to higher (weaker) geomagnetic activity in Northern Hemisphere (NH) winter for $B_y\! >\! 0$ ($B_y\! <\! 0$).  
In this paper we quantify the size of this explicit $B_y$-effect with respect to the solar wind coupling function, both for Northern and Southern high-latitude geomagnetic activity.  
We show that high-latitude geomagnetic activity is significantly (by about 40-50\%) suppressed for $B_y\!<\!0$ in NH winter and for $B_y\!>\!0$ in SH winter. 
When averaged over all months, high-latitude geomagnetic activity in NH is about 12\% weaker for $B_y\!<\!0$ than for $B_y\!>\!0$. 
The $B_y$-effect affects the westward electrojet strongly but hardly at all the eastward electrojet. 
We also show that the suppression of the westward electrojet in NH during $B_y\!<\!0$ maximizes when the Earth's dipole axis points towards the night sector, i.e., when the auroral region is maximally in darkness. 

\end{abstract}

%
%

%



%
%
%

\section{Introduction}

The interaction of the solar wind and the interplanetary magnetic field (IMF) with the terrestrial magnetic field generates geomagnetic activity and various other phenomena in the near-Earth space.
One of the main goals of solar-terrestrial physics is to understand the details of the different physical processes involved in this interaction.  
A better theoretical understanding of this interaction will allow, e.g., for a better prediction of geomagnetic activity and related space weather hazards, such as the charging and loss of satellites and geomagnetically induced currents in power lines.  

The most important parameter for solar wind-magnetosphere coupling is the north-south ($B_z$) component of the IMF in the geocentric solar magnetospheric (GSM) coordinate system, which controls magnetic reconnection at the subsolar magnetopause \citep{Dungey_1961}.
Accordingly, IMF $B_z$ is the key parameter also for geomagnetic activity, and is included in different solar wind-magnetosphere coupling functions, such as the Kan-Lee electric field $E_{KL} = vB_T^2\sin^2(\theta/2)$ \citep{Kan_Lee_1979} and the Newell universal coupling function $d\Phi_{MP}/dt = v^{4/3}B_T^{2/3}\sin^{8/3}(\theta/2)$ \citep{Newell_2007}.
In these expressions $v$ is solar wind speed, $B_T = \sqrt{B_z^2 + B_y^2}$ and $\theta = \arctan(B_y/B_z)$ is the so-called clock angle.
The same clock-angle dependence as in $E_{KL}$ also appears in the recently developed Borovsky coupling function \citep{Borovsky_2014}. 
Note that IMF $B_y$ is included in these coupling functions, but its effect is independent on its polarity (sign), due to the symmetry of factors appearing in $B_T$ and $\theta$. 
In this paper we use the Newell universal coupling function because it is optimized for high-latitude geomagnetic indices, such as the $AL$ index \citep{Davis_1966}, which primarily measures the westward electrojet in the Northern Hemisphere (NH).
However, the main results of this paper do not depend on the choice of the coupling function.

While the polarity of IMF $B_y$ does not have any independent role in the solar wind-magnetosphere coupling functions, it plays a significant role in modulating the IMF $B_z$-component observed in the GSM coordinate system via the Russell-McPherron (RMP) effect \citep{Russell_1973}.
The Russell-McPherron effect arises due to the seasonally (and diurnally) changing angle between the solar equatorial plane and the GSM z-axis.  
During spring (fall) an equatorial IMF vector pointing toward (away from) the Sun has a southward $B_z$-component in the GSM coordinate system, which enhances geomagnetic activity at this time.
This effect is included in the solar wind-magnetosphere coupling functions. 
Note also that the RMP effect maximizes on April 5 and October 5, i.e., the maximum effect is shifted from the equinoxes toward the following solstices.

There are also some magnetospheric and ionospheric phenomena for which the polarity of IMF $B_y$ plays an independent role. 
For example, \citet{Svalgaard_1968} and \citet{Mansurov_1969} showed that the daily variation of the magnetic field at high latitudes depends on the IMF sector polarity.
\citet{Friis-Christensen_1972} showed that this Svalgaard-Mansurov effect is due to the $B_y$-component of the IMF ($B_x$ being insignificant).
Later studies using ground-based magnetic field observations \citep{Friis-Christensen_1985} and radar measurements \citep{Ruohoniemi_1996, Ruohoniemi_2005, Pettigrew_2010} have shown that IMF $B_y$ controls the shape of polar cap convection pattern and the amplitude of the cross-polar cap potential.

Recently, \citet{Laundal_2016} and \citet{Friis-Christensen_2017} showed that there is a seasonally dependent effect of the IMF $B_y$ polarity in the $AL$ index.
They found that in NH winter (NH summer) $|AL|$ is greater (smaller) under $B_y > 0$ than under $B_y < 0$.
This is partly supported by \citet{Smith_2017}, who showed that the auroral electrojet currents (not differentiating westward or eastward electrojets), derived from observations of different polar-orbiting satellites, are enhanced in NH winter for $B_y > 0$ and in the southern hemisphere (SH) winter for $B_y < 0$. 
However, \citet{Smith_2017} did not find significant $B_y$ polarity effect in the summer hemisphere. 

In this paper we perform a detailed study on the effect of IMF $B_y$ to the high-latitude geomagnetic activity using geomagnetic indices from both hemispheres.
We will show that the Russell-McPherron effect can lead to a significant bias in any statistical studies quantifying the effect of $B_y$, if not properly accounted for. 
We will show that there is a strong, seasonally varying \textit{explicit} $B_y$-dependence which is not due to the RMP effect and which is not included in the coupling functions that describe the interaction between solar wind and geomagnetic activity (but do include, e.g., the RMP effect). 
The paper is organized as follows. 
Section 2 gives details of different databases and geomagnetic indices  used in this paper. 
In Sections 3 and 4 we study the effect of IMF $B_y$ to the $AL$ and $AU$ indices, respectively. 
In Section 5 we study the universal time (UT) dependence of the $B_y$-effect, and in Section 6 the $B_y$-effect in the Southern Hemisphere using  the $K$-index of the Syowa station.
In Section 7 we study possible biases to our results caused by IMF $B_x$-component.
Finally, we discuss our results and give our conclusions in Section 8.

\section{Data}

In this paper we use the hourly mean values of solar wind speed and the different IMF components in 1966-2015 from the OMNI2 database (\texttt{https://omniweb.gsfc.nasa.gov/}) time-shifted to the Earth's orbit, and the hourly $AL$ and $AU$ indices in 1966-2015 as proxies of high-latitude geomagnetic activity in the Northern Hemisphere.
The $AL$ and $AU$ indices are defined as the momentarily lowest ($AL$) and highest ($AU$) deviations in the horizontal magnetic field measured by a network of twelve stations at geomagnetic latitudes ranging from $60^{\circ}$N to $71^{\circ}$N. 
The $AL$ and $AU$ indices are proxies for the intensities of the westward and eastward electrojets in the Northern Hemisphere, respectively.

Due to the small number of long-running magnetic stations at southern high latitudes, there are no equivalents of $AL$ or $AU$ indices available for the Southern Hemisphere. 
In this paper we use the geomagnetic $K$-index measured at the Japanese Syowa station in 1966-2015 (geographic coordinates $69.0^{\circ}$S, $39.5^{\circ}$E; corrected geomagnetic coordinates: $65.6^{\circ}$S, $118^{\circ}$E). 
This is the longest-running geomagnetic index measured at a site, which is located in the proximity of the southern auroral electrojets.

\section{$AL$ index and solar wind coupling functions for positive and negative $B_y$}

\begin{figure}
\centering
\includegraphics[width=.9\linewidth]{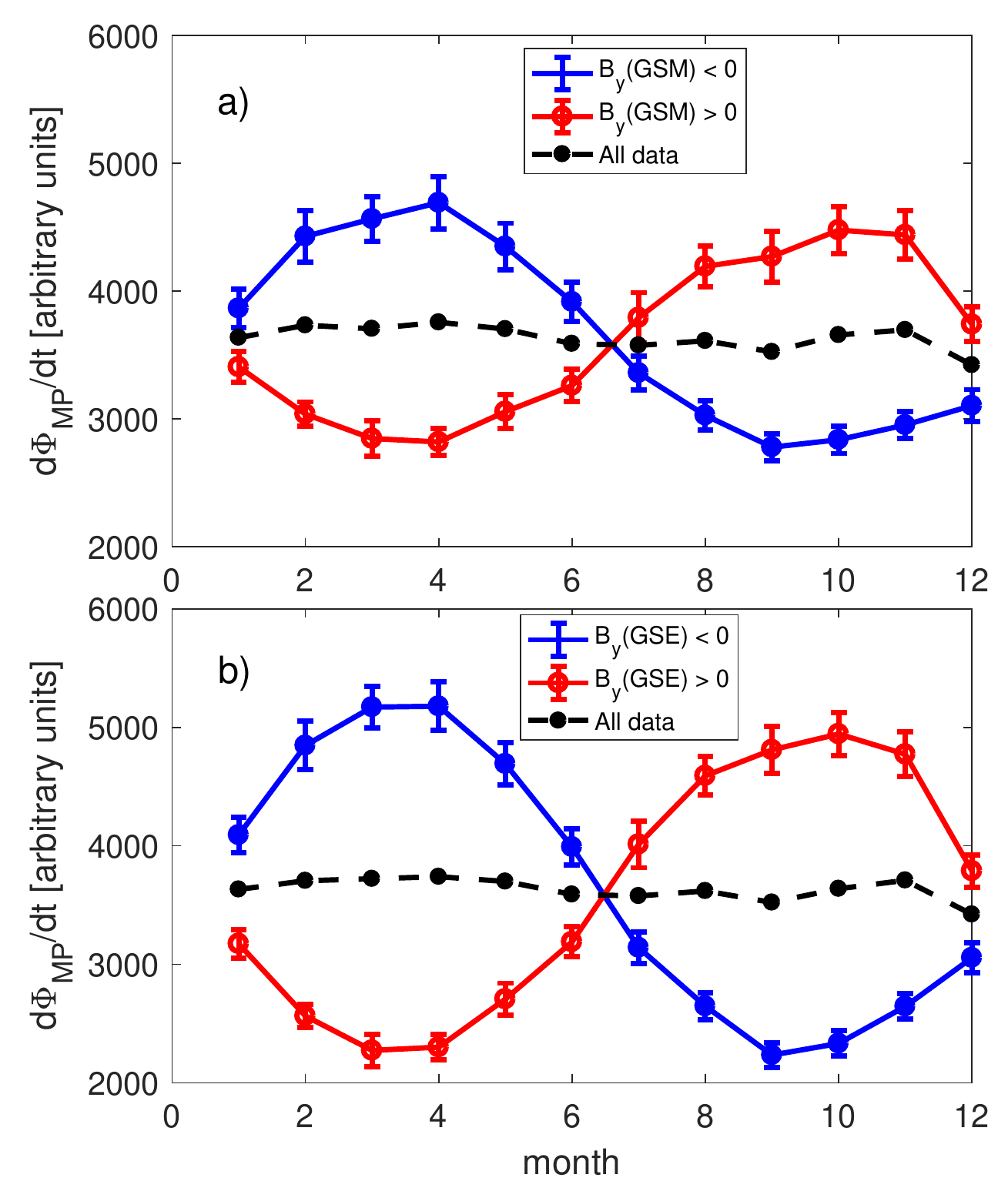}
\caption{Superposed monthly averages of the Newell universal coupling function $d\Phi_{MP}/dt$ in GSM coordinate system for the two polarities of IMF $B_y$. The polarity of IMF $B_y$ is defined in GSM coordinates in panel a) and in GSE coordinates in panel b). Standard errors of the superposed monthly averages are denoted by vertical bars.}
\label{newell_sector_means}
\end{figure}

Figure \ref{newell_sector_means} shows the superposed monthly averages of the Newell universal coupling function $d\Phi_{MP}/dt$ separately for $B_y > 0$ (away from the Sun) and $B_y < 0$ (toward the Sun) conditions in 1966-2015.
The polarity of $B_y$ is defined in GSM coordinates in Fig. \ref{newell_sector_means}a and in GSE coordinates in Fig. \ref{newell_sector_means}b. 
However, $d\Phi_{MP}/dt$ is calculated in the GSM coordinates in both Fig. \ref{newell_sector_means}a and \ref{newell_sector_means}b. 
Figure \ref{newell_sector_means} verifies the well-known, $B_y$-dependent seasonal variation, with maxima in $d\Phi_{MP}/dt$ in spring for $B_y<0$ and in fall for $B_y>0$ conditions.
This is in agreement with the Russell-McPherron effect, according to which, a toward (away) oriented field line attains an enhanced southward component in the GSM frame in spring (fall).
Because a typical IMF field line lies close to the ecliptic plane ($xy$-plane in GSE coordinates), defining the sign of $B_y$ in GSE coordinates yields a stronger seasonal variation in Fig. \ref{newell_sector_means}b than in Fig. \ref{newell_sector_means}a.

\begin{figure}
\centering
\includegraphics[width=\linewidth]{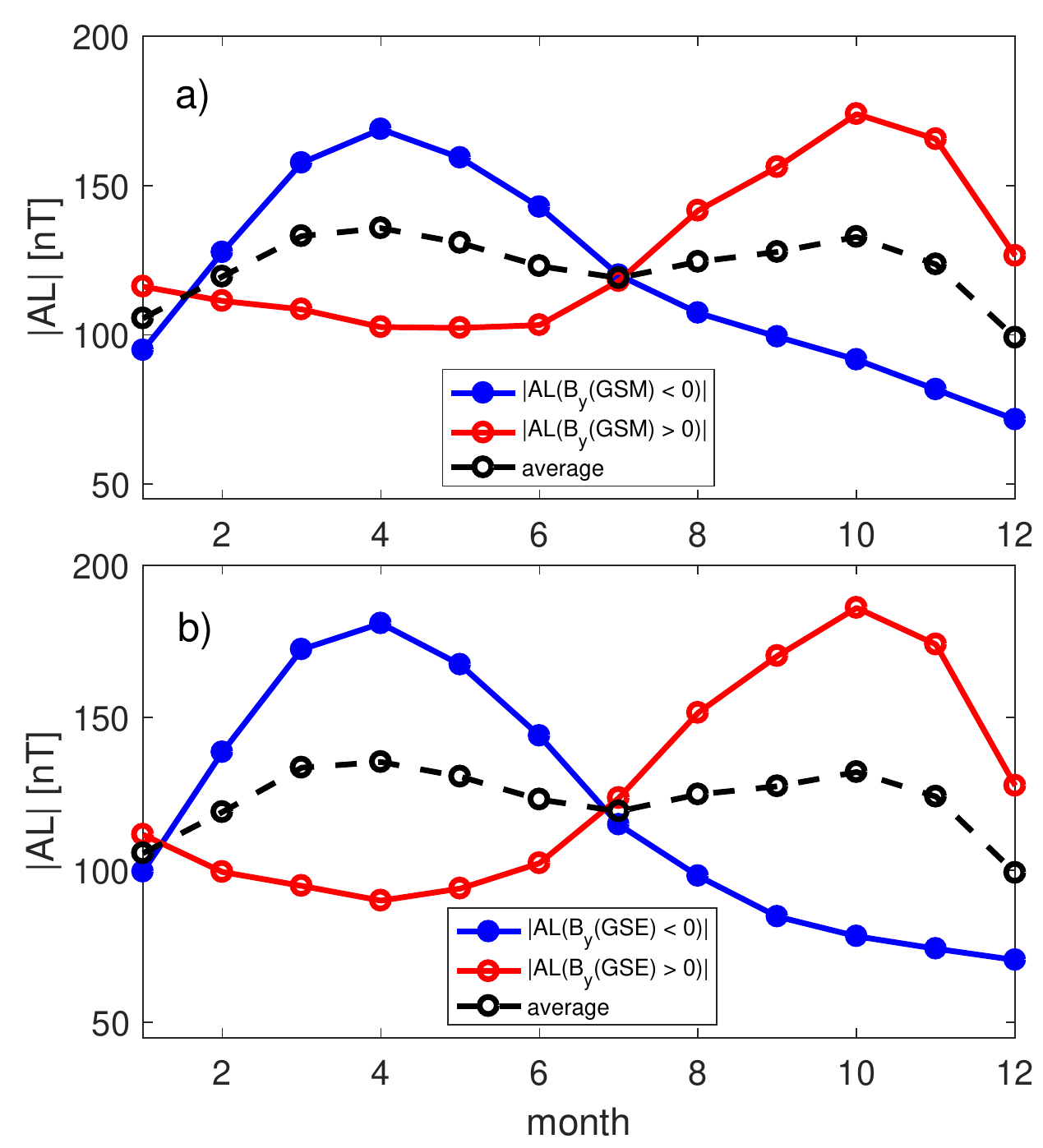}
\caption{Superposed monthly averages of the $|AL|$ index for the two polarities of IMF $B_y$. The polarity of IMF $B_y$ is defined in GSM coordinates in panel a) and in GSE coordinates in panel b). }
\label{AL_sector_means}
\end{figure}

Figure \ref{AL_sector_means} shows the superposed monthly averages of the $|AL|$ index for the two $B_y$ polarities, with the sector division made in the two coordinate systems. 
As the solar wind driver $d\Phi_{MP}/dt$, also the $AL$ index exhibits maxima in spring for $B_y<0$ and in fall for $B_y>0$, with the same peak months (April and in October) as in Fig. \ref{newell_sector_means}.
As noted above, April and October are the months of the maximum effect of the RMP mechanism.

There are, however, significant differences between the seasonal patterns of $d\Phi_{MP}/dt$ and the $AL$ index.
While the peaks and, especially, the minima of $d\Phi_{MP}/dt$ (Figure \ref{newell_sector_means}) are roughly equal for the two polarities of $B_y$, the minimum of the $|AL|$ index (Figure \ref{AL_sector_means}) in winter for $B_y<0$  is much lower than the minimum in spring/summer for $B_y>0$.
There are actually five consecutive months (September-January) during which $|AL(B_y<0)|$ is below any of the superposed monthly values of $|AL(B_y>0)|$.    
Thus, the fall-winter response of $|AL|$ to solar wind driving for $B_y<0$ conditions is considerably weaker than expected from the seasonal distribution of the solar wind driver function.

\begin{figure}
\centering
\includegraphics[width=.9\linewidth]{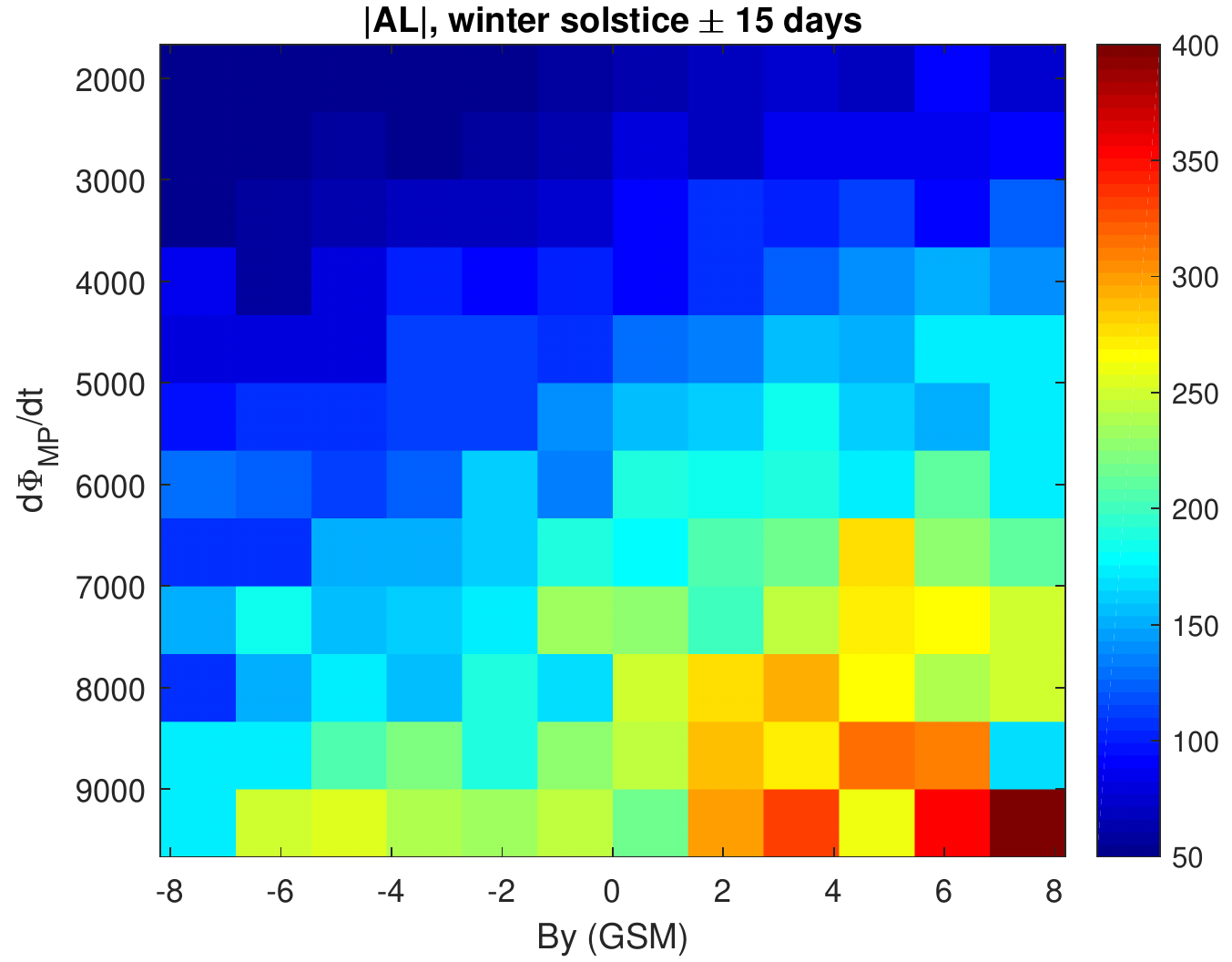}
\caption{Averages of $|AL|$ in 1966-2015 (in color code) during different values of $d\Phi_{MP}/dt$ and $B_y$. Only data within $\pm 15$ days from the NH winter solstice (Dec 21) are included.}
\label{AL_colorplot_winter}
\end{figure}

\begin{figure}
\centering
\includegraphics[width=.9\linewidth]{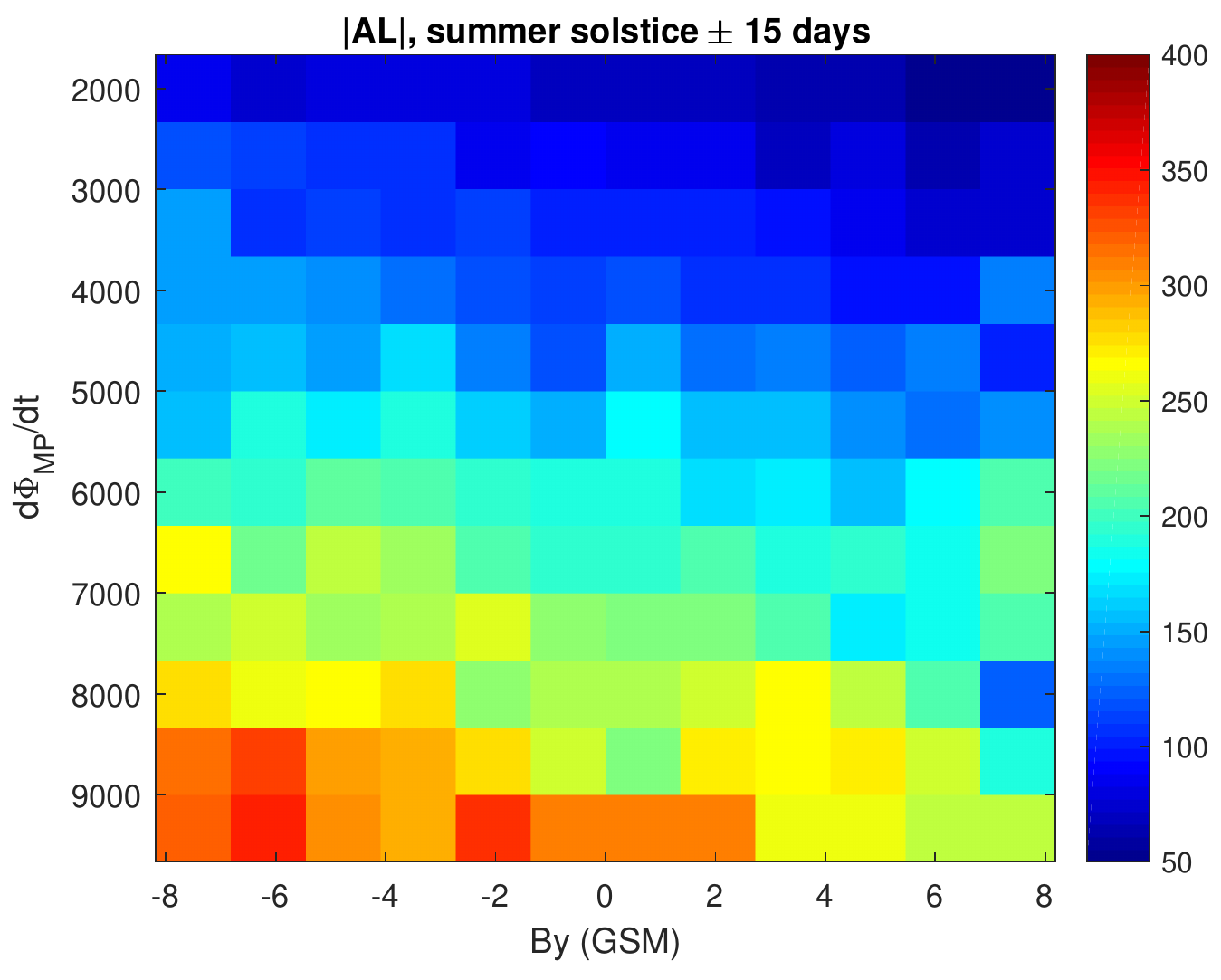}
\caption{Averages of $|AL|$ in 1966-2015 (in color code) during different values of $d\Phi_{MP}/dt$ and $B_y$. Only data within $\pm 15$ days from the NH summer solstice (Jun 21) are included.}
\label{AL_colorplot_summer}
\end{figure}

Because the seasonal patterns in Figures \ref{newell_sector_means} and \ref{AL_sector_means} are primarily due to the Russell-McPherron effect, Figure \ref{AL_sector_means} includes the RMP modulation of the strength of solar wind driving via the $B_y$-symmetric clock angle.    
To separate the possible explicit $B_y$-effect on the $AL$ index, we study the response of $AL$ to $B_y$ during given values of the solar wind driver function.
Figure \ref{AL_colorplot_winter} shows the average values of $|AL|$ in 1966-2015 for different measured values of $B_y$(GSM) and $d\Phi_{MP}/dt$ around winter solstice (December 21 $\pm 15$ days).
Figure \ref{AL_colorplot_winter} shows a clear asymmetry in the response of the $AL$ index to $B_y$(GSM): for a given value of $d\Phi_{MP}/dt$, $|AL|$ increases with increasing $B_y$(GSM). 
An opposite, but slightly weaker $B_y$-dependence can be seen in Figure \ref{AL_colorplot_summer} around the summer solstice (June 21 $\pm 15$ days).
Thus, there is an explicit $B_y$-dependence in $|AL|$, which suppresses $|AL|$ for $B_y<0$ in NH winter and for $B_y>0$ in NH summer.

\begin{figure}
\centering
\includegraphics[width=.9\linewidth]{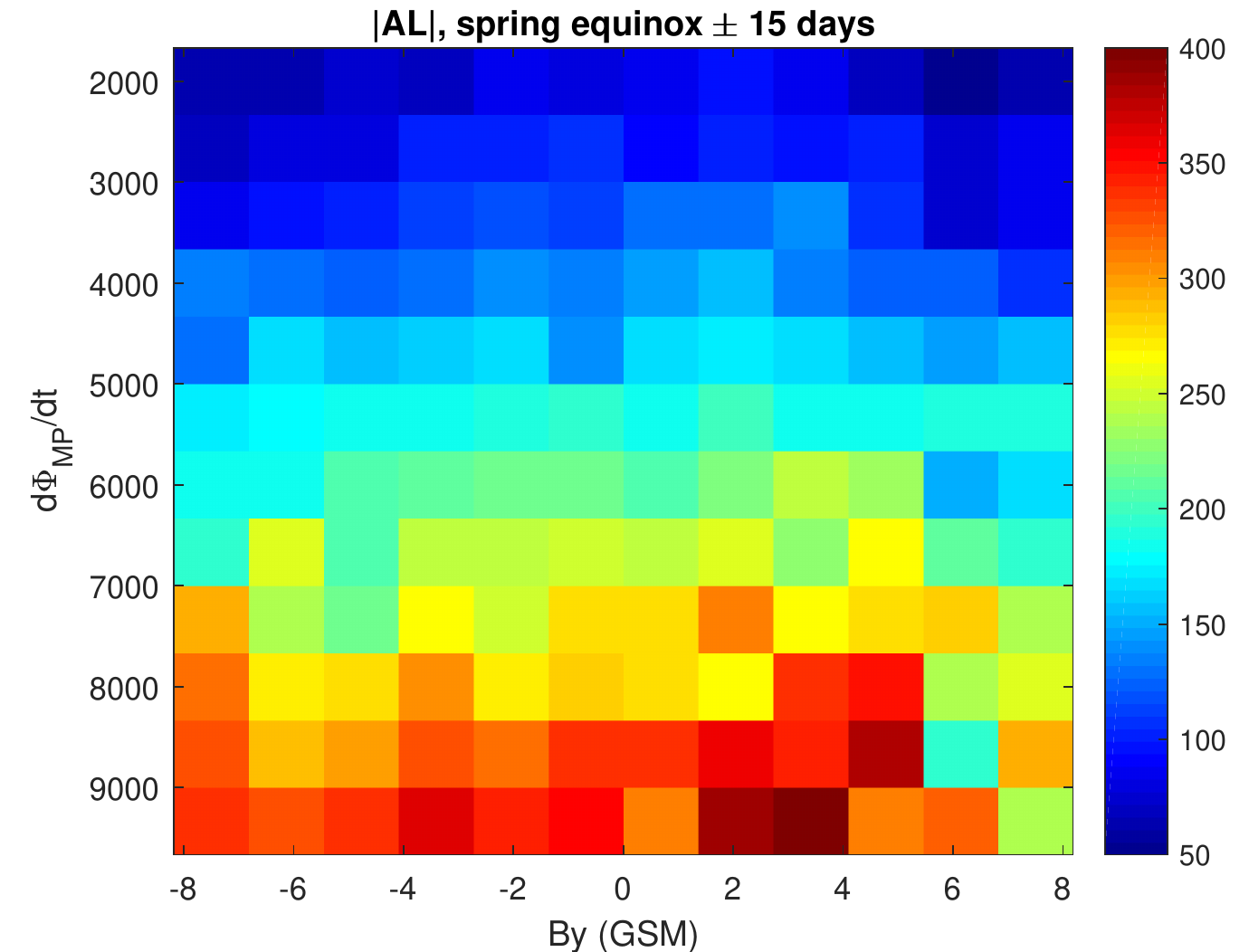}
\caption{Averages of $|AL|$ in 1966-2015 (in color code) during different values of $d\Phi_{MP}/dt$ and $B_y$. Only data within $\pm 15$ days from the NH spring equinox (Mar 20) are included.}
\label{AL_colorplot_spring}
\end{figure}

\begin{figure}
\centering
\includegraphics[width=.9\linewidth]{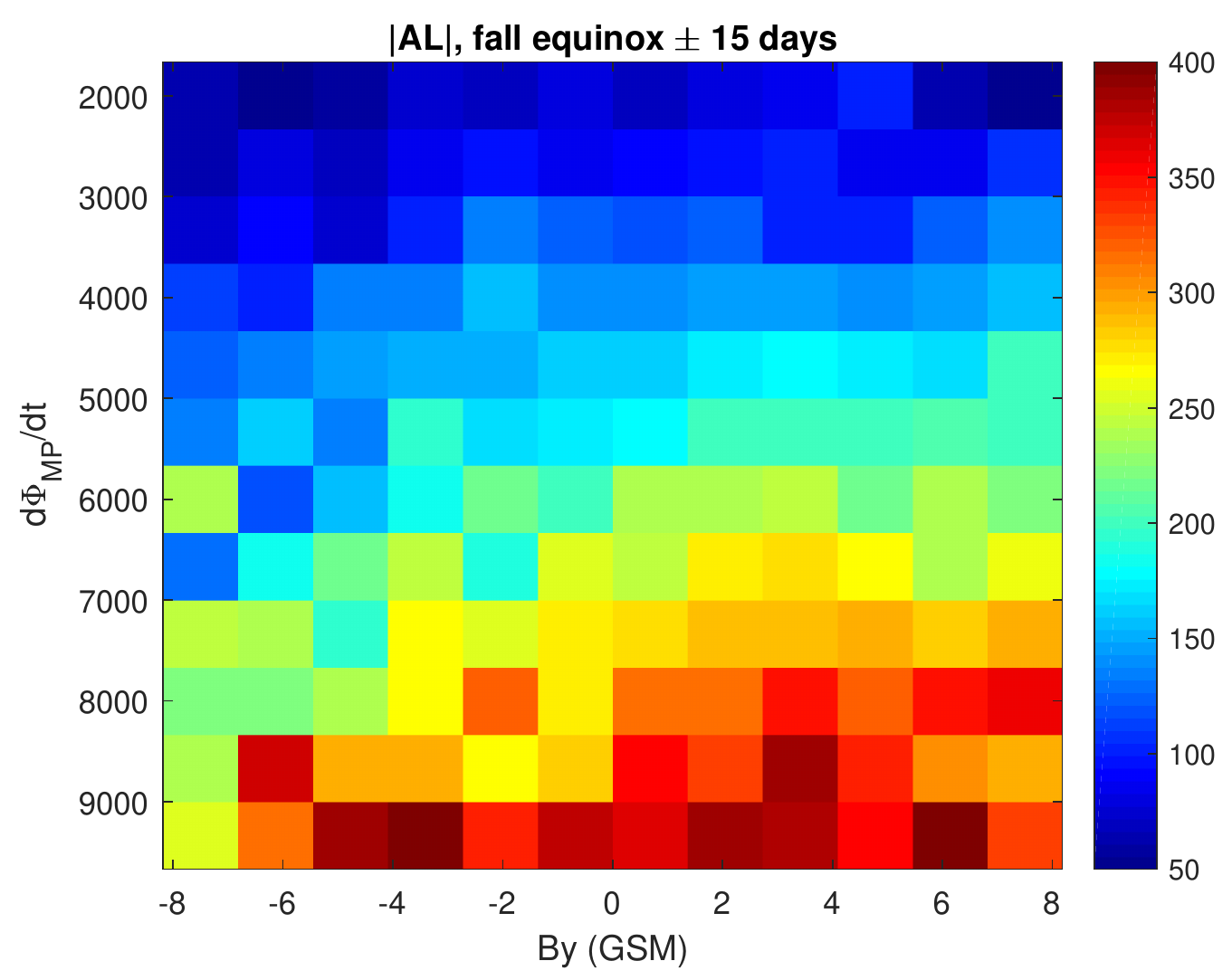}
\caption{Averages of $|AL|$ in 1966-2015 (in color code) during different values of $d\Phi_{MP}/dt$ and $B_y$. Only data within $\pm 15$ days from the NH fall equinox (Sep 22) are included.}
\label{AL_colorplot_fall}
\end{figure}

Figures \ref{AL_colorplot_spring} and \ref{AL_colorplot_fall} show the average  $|AL|$ as a function of $B_y$(GSM) and $d\Phi_{MP}/dt$ around spring and fall equinoxes (March 20 $\pm 15$ days and September 22 $\pm 15$ days, respectively). 
In spring the dependence of $|AL|$ for a given value of $d\Phi_{MP}/dt$ is quite symmetric with respect to the sign of $B_y$.
Only very large $B_y>0$ values lead to suppressed $|AL|.$ 
(Because this is only seen for one polarity of $B_y$, there is no saturation of $AL$ for large values of $|B_y|$). 
However, Figure \ref{AL_colorplot_fall} shows a weak but quite systematic  increase of $|AL|$ with $B_y$ in fall, in analogy to Fig. \ref{AL_colorplot_winter}.

\begin{figure}
\centering
\includegraphics[width=.9\linewidth]{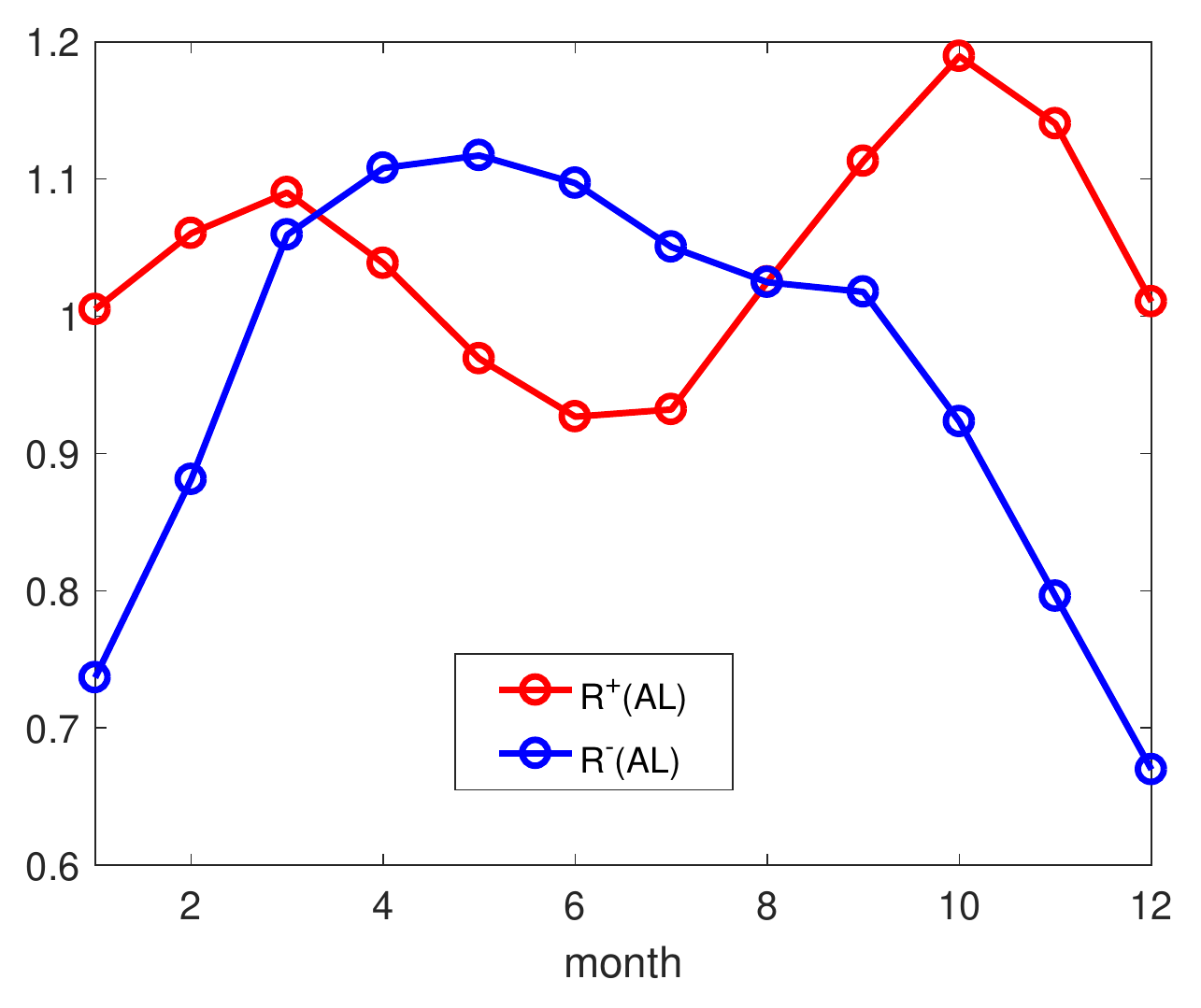}
\caption{Ratios of measured and predicted values of the $|AL|$ index for $B_y>0$ and $B_y<0$ ($R^{+}(AL)$ and $R^{-}(AL)$, respectively; see Eqs. 1 and 2)}.
\label{rats_separately}
\end{figure}

In principle, the $B_y$-effect seen, e.g., in winter (see Fig. \ref{AL_colorplot_winter}) might well be due either to the enhancement of the $|AL|$-index for $B_y>0$ or to the suppression of $|AL|$ for $B_y<0$.
To study this further, we show in Figure \ref{rats_separately} the ratios between the measured and predicted values of the $|AL|$ index
\begin{eqnarray}\label{rat_plus}
 &R^{+}(AL)& = \frac{|AL(B_y>0)|}{|a\cdot d\Phi_{MP}/dt(B_y>0) + b|}\\\label{rat_minus}
 &R^{-}(AL)& = \frac{|AL(B_y<0)|}{|a\cdot d\Phi_{MP}/dt(B_y<0) + b|}
\end{eqnarray}
where the coefficients $a=0.024$ nT$^{1/3}$/(km/s)$^{4/3}$ and $b = 12.3$ nT are obtained from the linear least squares fit using all hourly solar wind and $AL$ data in 1966-2015 without $B_y$-separation. 
Figure \ref{rats_separately} shows that the ratio $R^{-}(AL)$ is significantly below 1 in winter months, reaching the minimum of 0.67 in December. 
On the contrary, $R^{+} \approx 1$ in winter months. 
This proves that the explicit $B_y$-effect \textit{suppresses} geomagnetic activity for $B_y<0$ in winter rather than enhances it for $B_y>0$.     
The remaining semiannual variation in $R^{+}(AL)$ is probably mostly due to the so-called equinoctial effect \citep{Cliver_2000, Lyatsky_2001}, which modulates the relation between the solar wind driver and geomagnetic activity. 

\begin{figure}
\centering
\includegraphics[width=.9\linewidth]{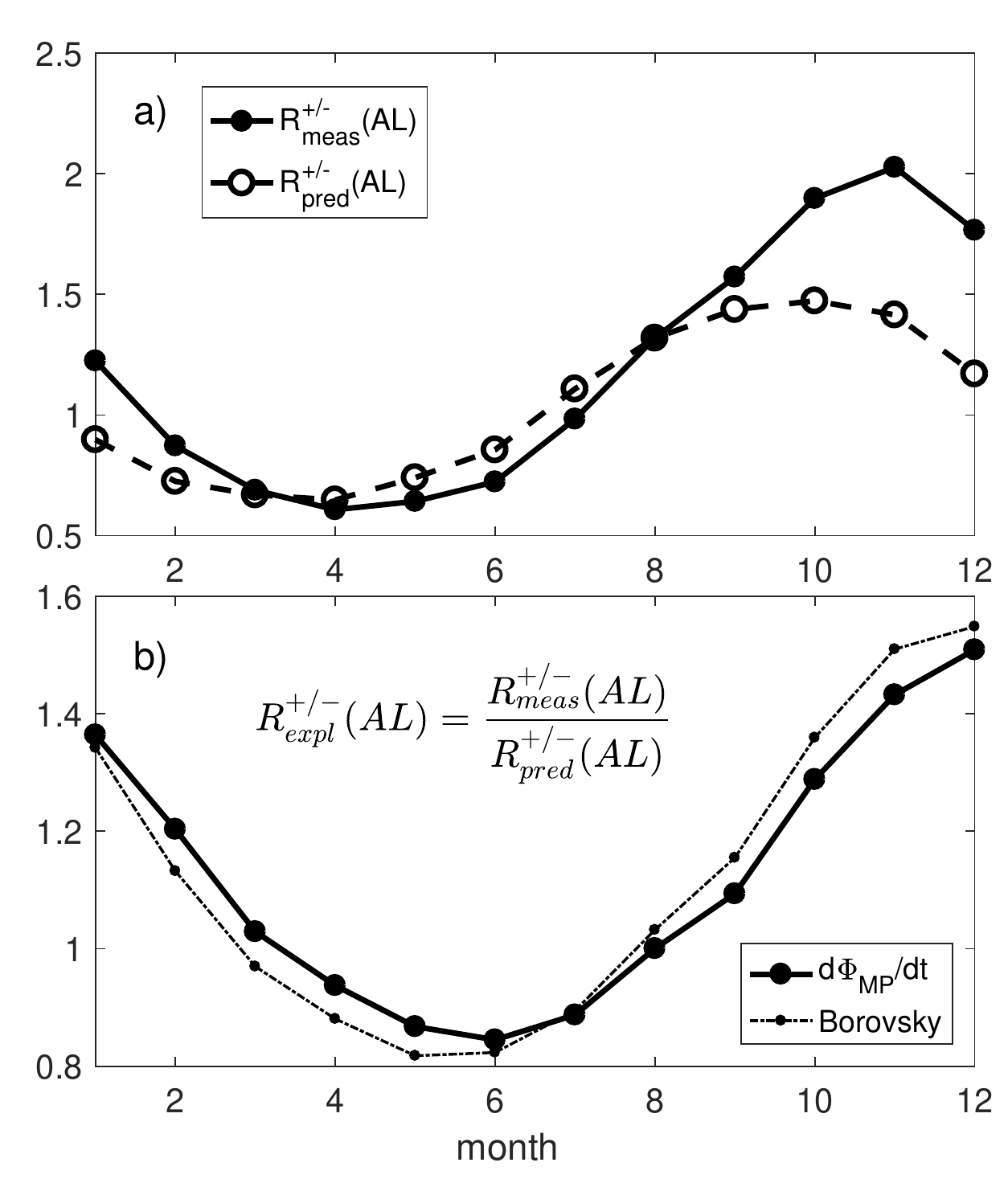}
\caption{a) Measured ($R^{+/-}_{meas}$) and predicted ($R^{+/-}_{pred}$) ratios of $|AL(B_y>0)|/|AL(B_y<0)|$ b) Ratio of measured and predicted ratios $R^{+/-}_{expl} = R^{+/-}_{meas}(AL)/R^{+/-}_{pred}(AL)$. Ratio $R^{+/-}_{expl}$ is calculated for $d\Phi_{MP}/dt$ and the Borovsky coupling function. }
\label{AL_meas_pred_ratio}
\end{figure}

To further quantify the strength of the explicit $B_y$-effect in the $AL$ index, we define the ratio  
\begin{equation}
R^{+/-}_{meas}(AL) = \frac{|AL(B_y>0)|}{|AL(B_y<0)|}
\end{equation}
calculated from the measured values of $|AL|$ and the corresponding ratio predicted from the solar wind driver function
\begin{equation}
R^{+/-}_{pred}(AL) = \frac{a\cdot d\Phi_{MP}/dt(B_y>0) + b}{a\cdot d\Phi_{MP}/dt(B_y<0) + b}.
\end{equation} 
These ratios are shown in Figure \ref{AL_meas_pred_ratio}a.
While $R^{+/-}_{meas}(AL)$ and $R^{+/-}_{pred}(AL)$ show qualitatively similar seasonal variations, there are some significant differences.
In particular, as expected from the comparison of Figures \ref{newell_sector_means} and \ref{AL_sector_means} and Figure \ref{rats_separately}, $R^{+/-}_{meas}$ attains significantly higher values than $R^{+/-}_{pred}$ in October, November, December and January. 
Interestingly, $R^{+/-}_{meas} > 1$ even in January, when the Russell-McPherron effect already favors $B_y < 0$, leading to $R^{+/-}_{pred} < 1$.
This strongly implies that the winter minimum of $|AL(B_y<0)|$ is a major effect in high-latitude geomagnetic activity, which is not due to the Russell-McPherron effect.

Figure \ref{AL_meas_pred_ratio}b shows the ratio of ratios  
\begin{equation}
R^{+/-}_{expl}(AL) = \frac{R^{+/-}_{meas}(AL)}{R^{+/-}_{pred}(AL)},
\end{equation}
which maximizes in winter and minimizes in summer.
For this plot we have included also the similar ratio $R^{+/-}_{expl}(AL)$ calculated using the Borovsky coupling function \citep{Borovsky_2014}, yielding a very similar result as $d\Phi_{MP}/dt$. 
This gives confidence that the results obatined in this paper are not limited to one specific coupling function. 
(Note also that even the clock angle dependencies are somewhat different in Newell and Borovsky functions).
The ratio $R^{+/-}_{expl}$ quantifies the strength of the explicit $B_y$-effect by removing not only the RMP effect but also other known causes of seasonal variation, like the equinoctial effect.
Note also that the maximum and the minimum of the ratio $R^{+/-}_{expl}(AL)$ occur exactly at summer and winter solstices.  
This (together with Figure \ref{rats_separately}) indicates that the response of the westward electrojet (of the Northern Hemisphere) to solar wind driving is considerably weaker in winter but slightly stronger in summer for $B_y<0$ than for $B_y>0$.
Averaging $R^{+/-}_{expl}(AL)$ over all 12 months yields to the overall average of 1.12.
Thus, the overall annual response of the westward electrojet to solar wind driving is 12\% weaker for $B_y<0$ than for $B_y>0$.
During the winter months (Nov-Jan) the ratio $R^{+/-}_{expl}(AL)$ is 1.44, indicating a highly significant effect.

\section{$AU$ index for positive and negative $B_y$}

\begin{figure}
\centering
\includegraphics[width=.9\linewidth]{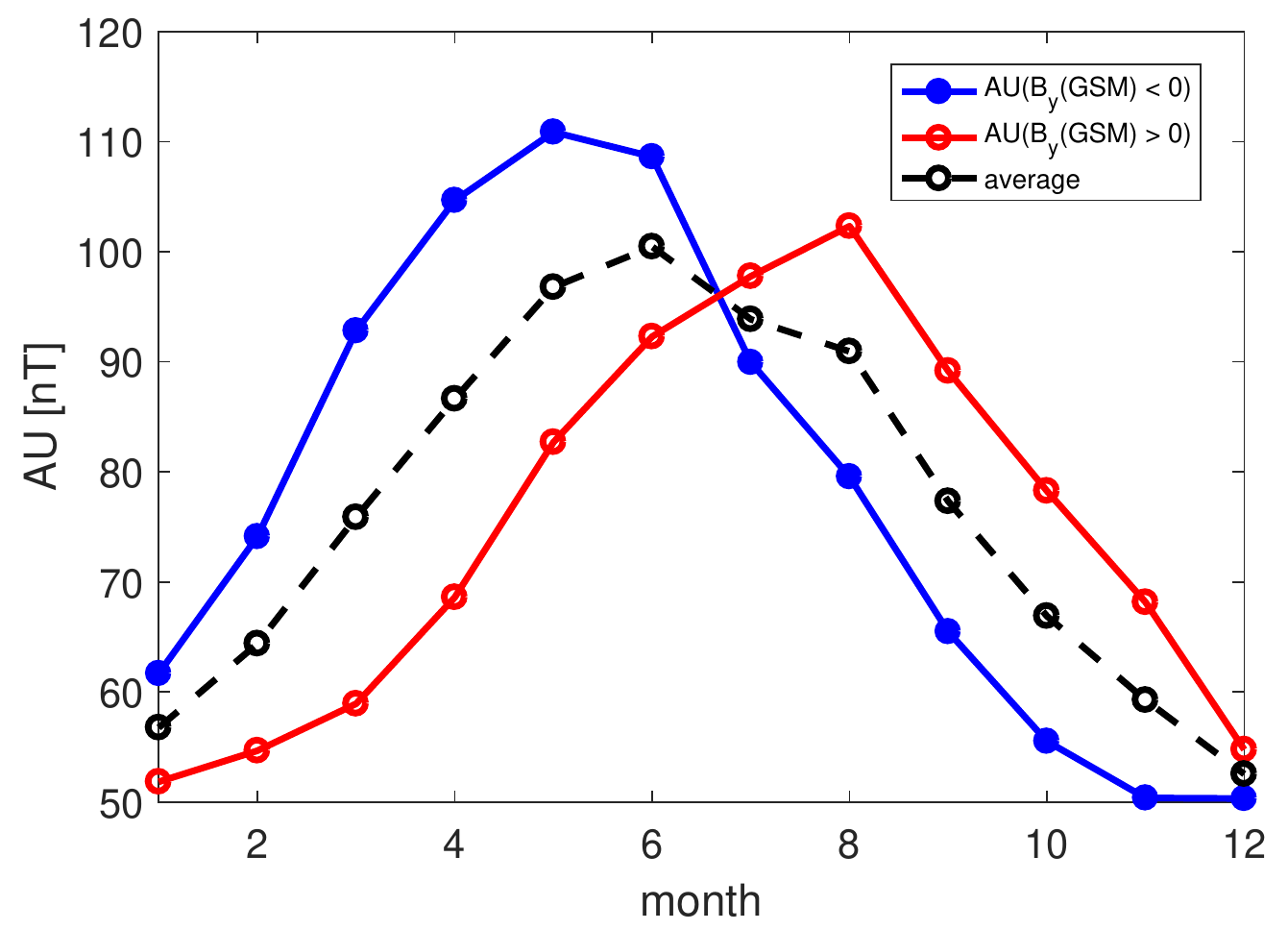}
\caption{Averages of the $AU$ index under under different polarities of IMF $B_y$ as a function of month.}
\label{AU_sector_means}
\end{figure}

Figure \ref{AU_sector_means}  shows the superposed monthly averages of the $AU$ index for the two IMF $B_y$ polarities in 1966-2015. 
The $AU$ index shows a very strong annual (summer-winter) variation, related to varying illumination of the ionosphere [see, e.g., \citet{Finch_2008}] which can be seen in the overall average of $AU$ (also included in Figure \ref{AU_sector_means}). 
The seasonal variation of illumination strongly affects $AU$ because the intensity of eastward electrojet maximizes at the afternoon sector, where ionospheric conductivity is dominated by solar EUV radiation.
Figure \ref{AU_sector_means} shows that the IMF $B_y$-component shifts the annual maximum of $AU$ to May for $B_y < 0$ and to August for $B_y > 0$, i.e., always towards the corresponding RMP month (April and October, respectively).

\begin{figure}
\centering
\includegraphics[width=.9\linewidth]{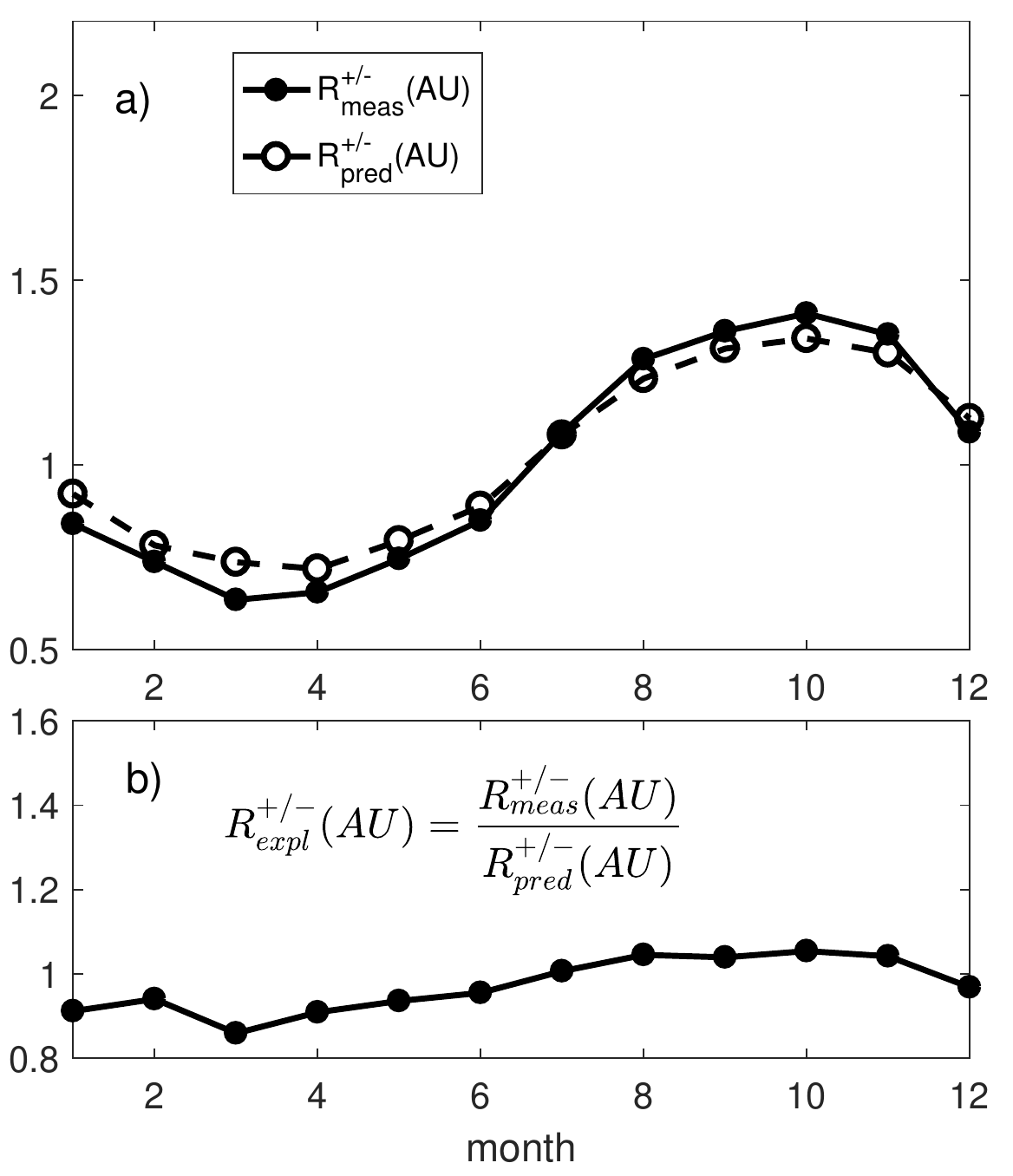}
\caption{a) Measured and predicted ratios of $AU(B_y>0)/AU(B_y<0)$ ($R^{+/-}_{meas}(AU)$ and $R^{+/-}_{pred}(AU)$, respectively) b) Ratio of measured and predicted ratios $R^{+/-}_{expl}(AU) = R^{+/-}_{meas}(AU)/R^{+/-}_{pred}(AU)$. }
\label{AU_meas_pred_ratio}
\end{figure}

Figure \ref{AU_meas_pred_ratio}a shows the measured ratio $R^{+/-}_{meas}(AU) = AU(B_y>0)/AU(B_y<0)$ and the corresponding predicted ratio $R^{+/-}_{pred}(AU)$ calculated in the same way as for the $AL$ index above (now $a=0.0129$ nT$^{1/3}$/(km/s)$^{4/3}$ and $b = 29.7$ nT).
Unlike for the $AL$ index, $R^{+/-}_{meas}(AU)$ and $R^{+/-}_{pred}(AU)$ are very similar and their ratio $R^{+/-}_{expl}(AU)$ depicted in Figure \ref{AU_meas_pred_ratio}b remains close to one for all months.
(The overall mean of $R^{+/-}_{expl}(AU)$ is 0.97).
Note how closely Figure \ref{AU_meas_pred_ratio}a reproduces the seasonal pattern of the Russell-McPherron effect.
This proves that the Russell-McPherron effect plays almost an exclusive role in varying the seasonal variation of the $AU$ index with IMF $B_y$ polarity (Figure \ref{AU_sector_means}).
Thus, the $AU$ index does not have any notable \textit{explicit} dependence on the IMF $B_y$-component beyond the Russell-McPherron effect.
Thus, there is an \textit{explicit} $B_y$-effect only in the westward electrojet. 
	
\section{UT dependence of the explicit $B_y$ effect in $AL$}

\begin{figure}
\centering
\includegraphics[width=.9\linewidth]{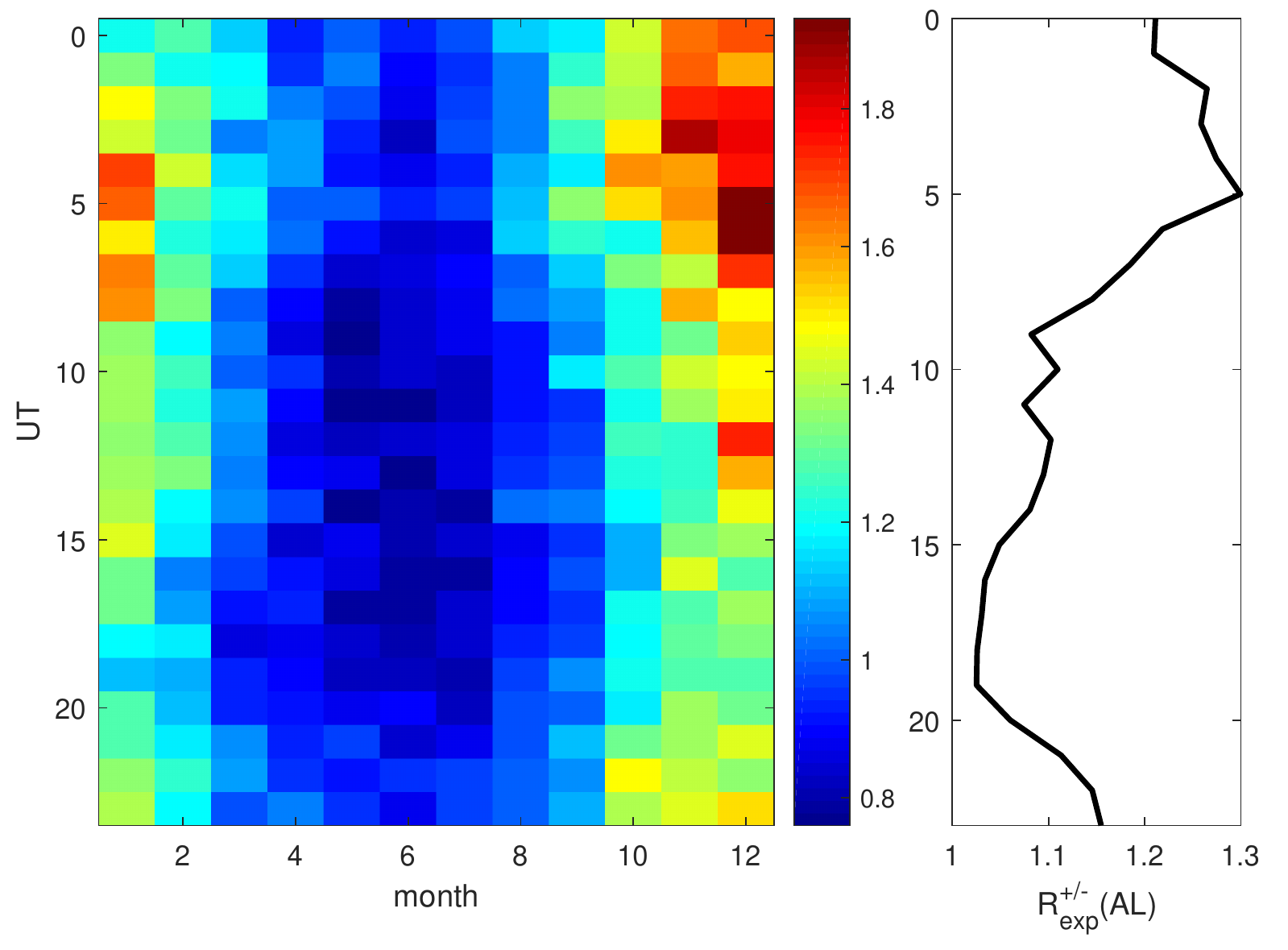}
\caption{Left: Ratio $R^{+/-}_{expl}(AL)$ calculated separately in different UT hours and months. Right: $R^{+/-}_{expl}(AL)$ in different UT hours averaged over all months. }
\label{AL_meas_pred_ratio_UT}
\end{figure}

Figure \ref{AL_meas_pred_ratio_UT} shows the ratio $R^{+/-}_{expl}(AL)$ in different months and different UT hours.
While the ratio shows qualitatively the same seasonal pattern for all UT hours as in Fig. \ref{AL_meas_pred_ratio}, the highest values are found around 5 UT and the lowest values approximately 12 hours later around 17-19 UT.
This can be best seen in the right panel, which shows the averages of $R^{+/-}_{expl}(AL)$ over all 12 months. 
Interestingly, at 5 UT the Earth's dipole axis points towards the night sector (anti-sunward direction) in the Northern Hemisphere, while the maximal tilt towards the noon (sunward direction) takes place at 17 UT.
This UT variation of $R^{+/-}_{expl}(AL)$, together with its seasonal variation discussed above (see Fig. \ref{AL_meas_pred_ratio}), strongly indicate that the explicit $B_y$-dependence in the $AL$ index is related to (lack of) illumination and is effective when the auroral region of the Northern Hemisphere is maximally in darkness.

\section{IMF $B_y$-effect in the Southern Hemisphere}

In order to study whether the explicit $B_y$-dependence also appears in the high-latitude geomagnetic activity of the Southern Hemisphere, we repeat the above analysis using the $K$-index of the Japanese Syowa station.  
Because the Syowa station is located close to the Southern auroral region, its $K$-index is primarily affected by the auroral electrojets. 
Since Syowa $K$-index is a local measure of geomagnetic activity, we cannot study the UT variation. 
Here we only use the two three-hour $K$-index bins (0-2 UT and 3-6 UT corresponding 21-23 LT and 0-2 LT) closest to the local midnight sector, where the effect of the westward electrojet is largest.

\begin{figure}
\centering
\includegraphics[width=.9\linewidth]{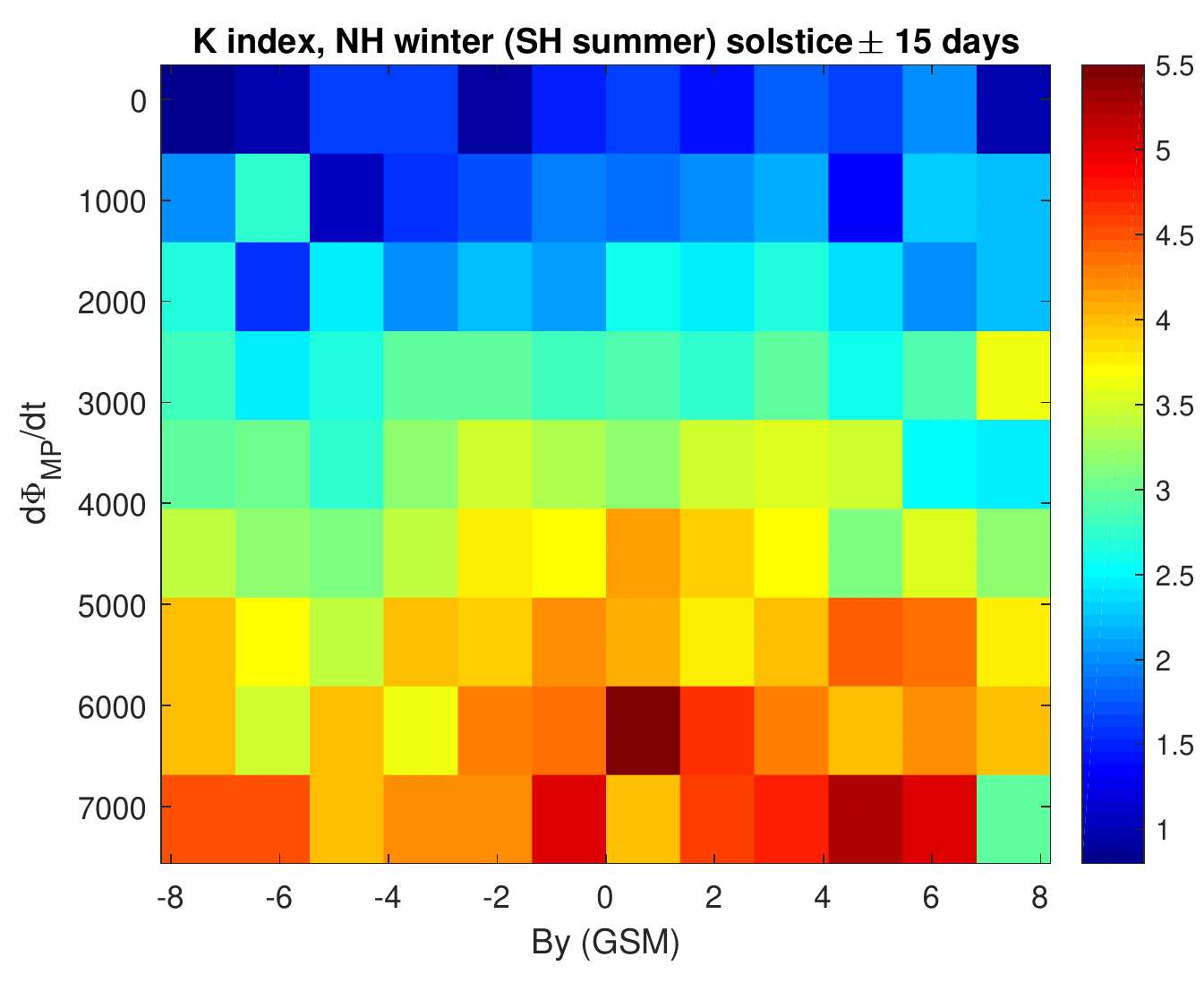}
\caption{Averages of the $K$-index of Syowa station in color code during different levels of $d\Phi_{MP}/dt$ and $B_y$ in NH winter (SH summer).}
\label{K_colorplot_winter}
\end{figure}
 
\begin{figure}
\centering
\includegraphics[width=.9\linewidth]{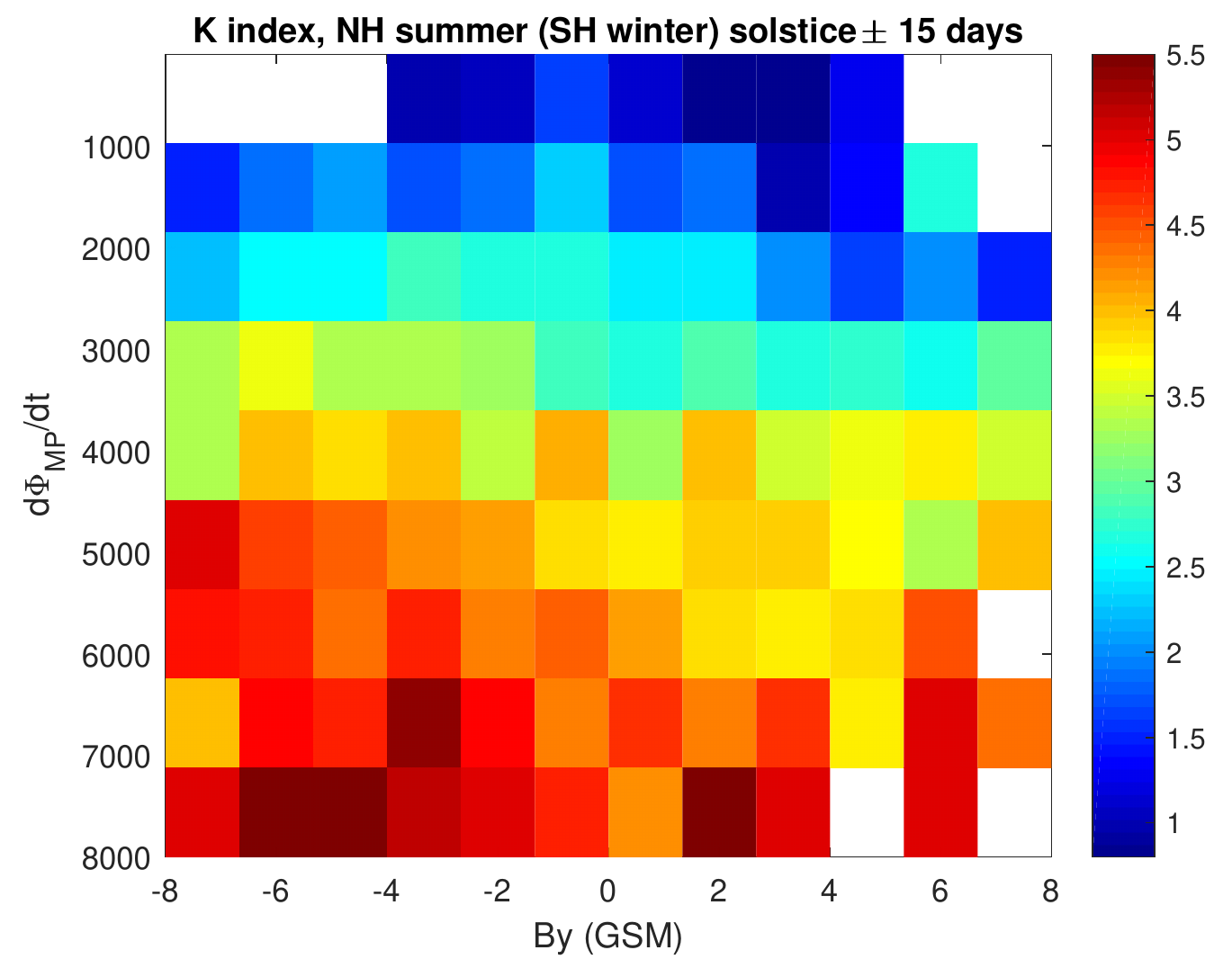}
\caption{Same as Figure \ref{K_colorplot_winter}, but for NH summer (SH winter).}
\label{K_colorplot_summer}
\end{figure}

\begin{figure}
\centering
\includegraphics[width=.9\linewidth]{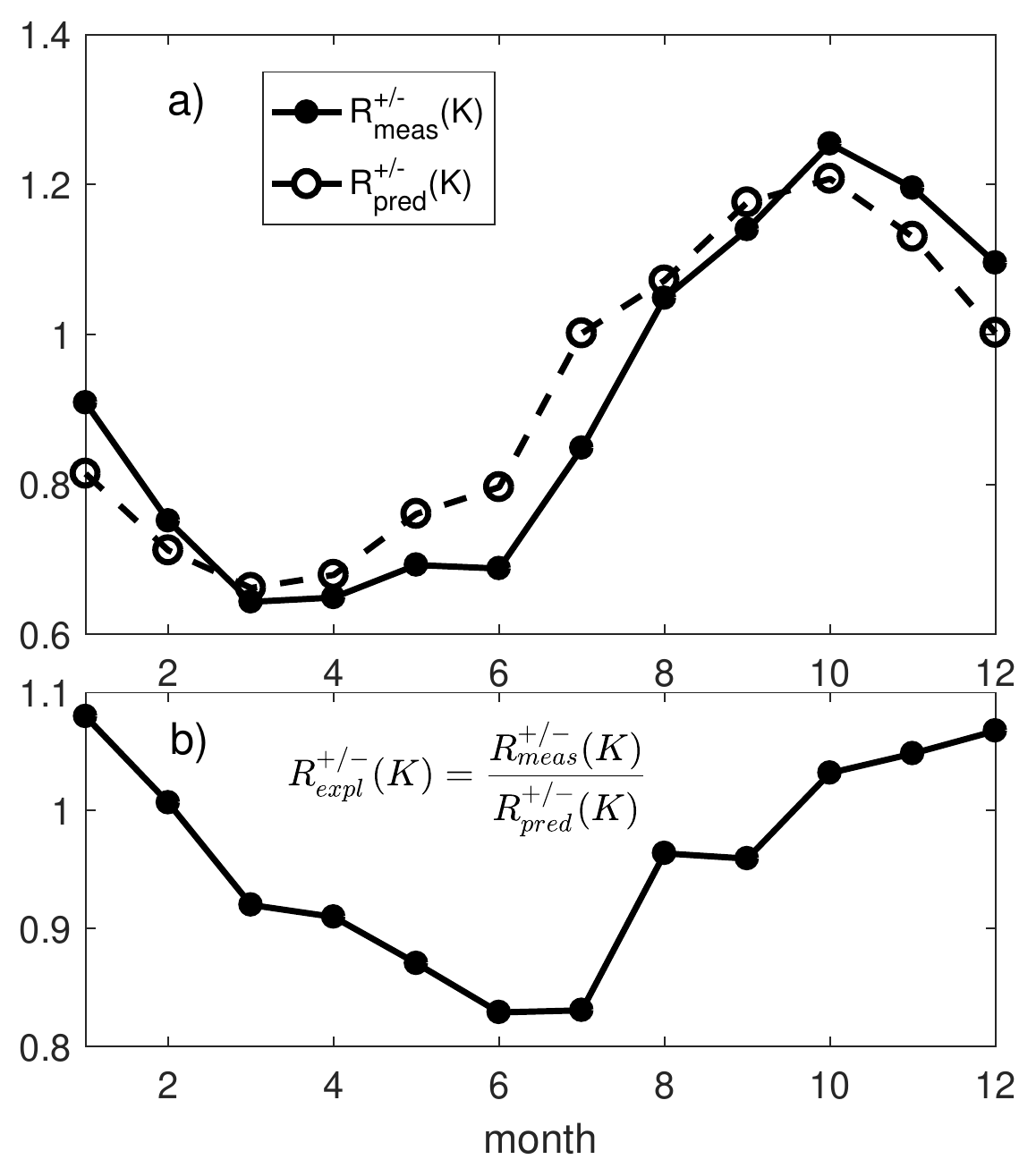}
\caption{a) Measured and predicted ratios of $K(B_y>0)/K(B_y<0)$ ($R^{+/-}_{meas}(K)$\\ and $R^{+/-}_{pred}(K)$, respectively). b) Ratio of measured and predicted ratios\\ $R^{+/-}_{expl}(K) = R^{+/-}_{meas}(K)/R^{+/-}_{pred}(K)$. }
\label{K_meas_pred_ratio}
\end{figure}

Figures \ref{K_colorplot_winter} and \ref{K_colorplot_summer} show the averages of the Syowa $K$-index for different values of $d\Phi_{MP}/dt$ and $B_y$ in Northern Hemisphere winter and summer, or SH summer and winter, respectively, in analogy with Figures \ref{AL_colorplot_winter} and \ref{AL_colorplot_summer} for the $|AL|$ index.
The stronger and more systematic dependence of the Syowa $K$-index on $B_y$ polarity is seen during NH summer (SH winter), when geomagnetic activity decreases with $B_y$ for a given $d\Phi_{MP}/dt$. 
A weaker $B_y$-dependence is seen during NH winter (SH summer). 
These effects are further quantified in Figure \ref{K_meas_pred_ratio}a, which shows the monthly ratios $R^{+/-}_{meas}(K)$ and $R^{+/-}_{pred}(K)$. 
The predicted values of the $K$-index are calculated from simultaneous three-hour averages of $d\Phi_{MP}/dt$ ($a=0.00029$ nT$^{-2/3}$/(km/s)$^{-4/3}$ and $b = 1.22$).
The explicit $B_y$-dependence of the Syowa $K$-index is clearly seen in Figure \ref{K_meas_pred_ratio}b, which shows the ratio $R^{+/-}_{expl}(K)$. 
The explicit $B_y$-effect maximizes during SH winter (in June), when the $K$-index is suppressed for $B_y > 0$ and $R^{+/-}_{expl}(K)$ = 0.81.
Thus, the suppression of high-latitude geomagnetic activity in local winter is due to $B_y>0$ in SH and $B_y<0$ in NH.
Figures \ref{K_colorplot_winter} and \ref{K_meas_pred_ratio}b also show that in SH summer $B_y<0$ leads to a slight suppression of the $K$-index and $R^{+/-}_{expl}(K) > 1$.  

The explicit $B_y$-effect is notably weaker in the (local) Syowa $K$-index than in the $AL$ index covering a range of longitudes in NH (compare Figures \ref{AL_meas_pred_ratio}b and \ref{K_meas_pred_ratio}b).  
The maximum $B_y$-effect in the Syowa $K$-index occurs in June, when $1/R^{+/-}_{expl}(K) \approx 1.22$, while the maximum of $R^{+/-}_{expl}(AL)$ in December is about 1.5.
Thus, $R^{+/-}_{meas}(K)$ is 22\% smaller than expected from the solar wind coupling function in June, while $R^{+/-}_{meas}(AL)$ is 50\% larger than expected in December.
This difference in the magnitude of the explicit $B_y$-effect is probably due to the UT variation of the $B_y$-effect, which in the SH is shifted by 12 hours from the UT the variation of NH.
Thus, the strongest $B_y$-effect for the Southern Hemisphere (around 17 UT ) is not observed in the local night sector of Syowa (UT 21-02).
Summarizing, there is an opposite explicit $B_y$-dependence in local winter in the Southern Hemisphere, with $B_y>0$ conditions leading to suppressed geomagnetic activity.

\section{Effect of IMF $B_x$}

\begin{figure}
\centering
\includegraphics[width=.9\linewidth]{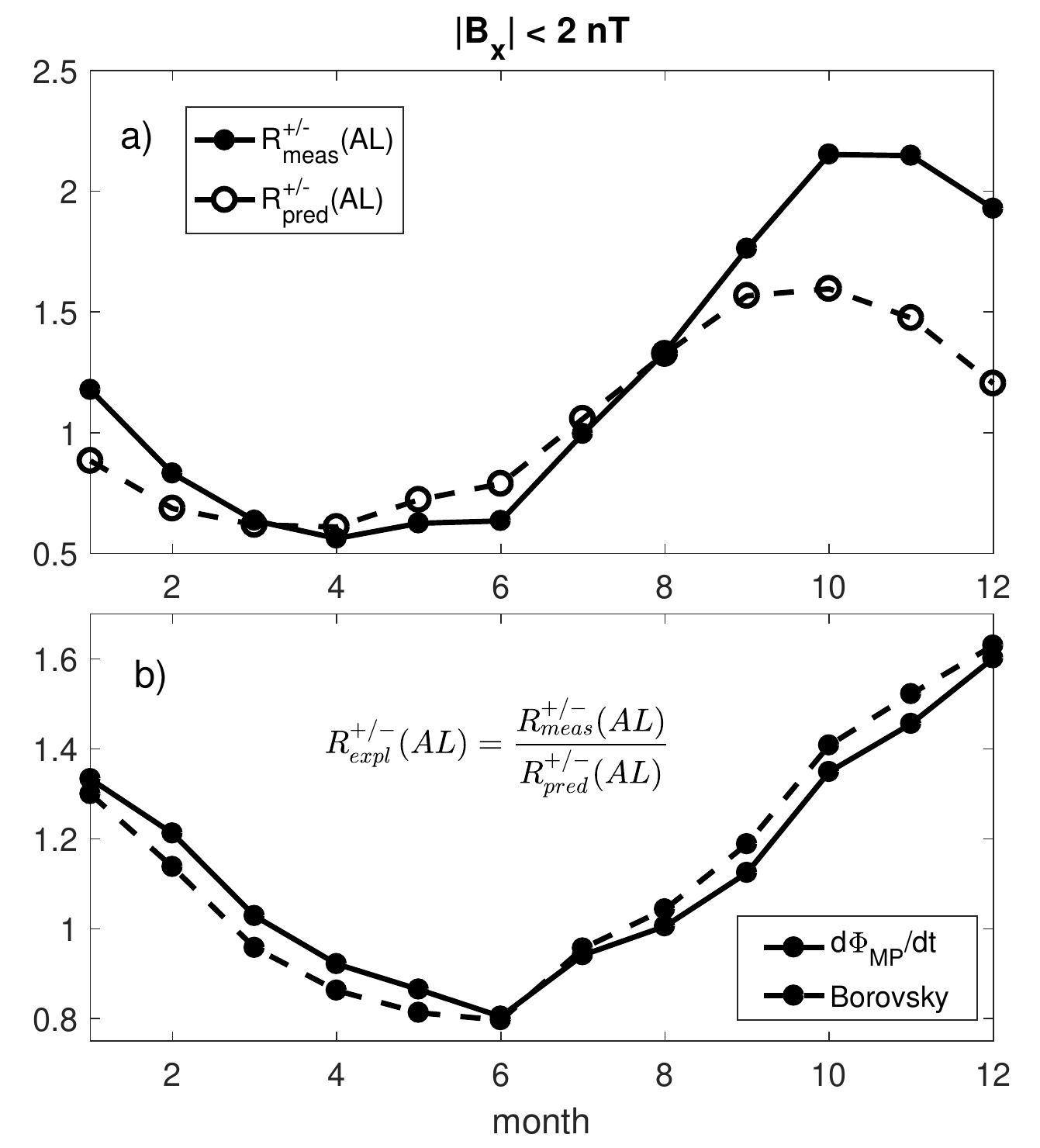}
\caption{Same as Figure \ref{AL_meas_pred_ratio}, but for $|B_x|<2$ nT.}
\label{AL_Bx}
\end{figure}

In earlier sections we have quantified the effect of IMF $B_y$ to high-latitude geomagnetic activity without considering a possible effect of IMF $B_x$. 
Because a typical IMF field line follows the Parker spiral, there is a well-known anticorrelation between $B_y$ and $B_x$. 
Thus, because $B_y$ and $B_x$ are not independent, the analysis in the earlier sections could be biased by $B_x$. 
To test whether effect of IMF $B_y$ dominates over $B_x$, in Figure \ref{AL_Bx} we repeat the above analysis depicted in Figure \ref{AL_meas_pred_ratio} under the additional constraint that $|B_x|<2$ nT. 
Because there are no significant differences between Figures \ref{AL_meas_pred_ratio} and \ref{AL_Bx}, we can conclude that the possible $B_x$-effect is much weaker that the $B_y$-effect.

We note that \citet{Laundal_2018} found that auroral currents are only weakly ($\leq 10$\%) affected by $B_x$.  
They suggested that the solar wind-magnetosphere coupling is more efficient when the tilt angle of the Earth's magnetic field and $B_x$ have the same sign, when the reconnection line moves towards subsolar magnetopause \citep{Hoilijoki_2014}, making reconnection more efficient.
This would lead to strongest geomagnetic activity for $B_x>0$ ($B_x<0$) in NH summer solstice at 17 UT (winter solstice at 5 UT).  
Thus, the $B_x$-effect should have similar seasonal/UT variation as the $B_y$-effect, making the separation of these two effects even more difficult. 
Detailed future studies are needed for more accurate quantification of $B_x$-effect and its physical mechanism.  

\section{Discussion and conclusions}

In this paper we have studied how the IMF $B_y$-component affects to high-latitude geomagnetic activity, using geomagnetic indices from both hemispheres.
We have confirmed the earlier observations (\citet{Laundal_2016}; \citet{Friis-Christensen_2017}; \citet{Smith_2017}) that the IMF $B_y$ polarity and amplitude modulate the strength of the westward electrojet so that the westward electrojet is weaker for $B_y < 0$ in NH winter and for $B_y>0$ in SH winter.
We have shown here that this explicit $B_y$-dependence is not due to the Russell-McPherron effect or other known effects in the solar wind-magnetosphere coupling (such as the equinoctial effect). 
We have also demonstrated that the explicit $B_y$-effect leads to suppression (for $B_y<0$) rather than enhancement (for $B_y>0$) of high-latitude geomagnetic activity in NH winter.

Furthermore, we have shown that the explicit $B_y$-effect depends strongly on UT. 
The strongest $B_y$ effect to the $AL$ index is observed at 5 UT in (NH) winter, when the Earth's dipole axis points towards the night.
This UT variation, together with the seasonal variation, verify that the explicit $B_y$-dependence of high-latitude geomagnetic activity maximizes when the local auroral region is maximally shadowed during local winter solstices.

\citet{Ruohoniemi_2005} and \citet{Pettigrew_2010} have found that the ionospheric convection in NH (measured by the cross-polar cap potential) is stronger in winter for $B_y > 0$ and in summer for $B_y < 0$. 
They also found that the IMF $B_y$-effect is especially strong in the dawn convection cell, which is connected to the westward electrojet. 
Thus, these studies are in agreement with the order of $B_y$-dependence and with our finding of a strong $B_y$-effect in the westward electrojet but not in the eastward electrojet.
\citet{Friis-Christensen_2017} suggested that $B_y$ modulates the intensity of the substorm current wedge, possibly explaining the $B_y$-dependence in the $AL$ index.  
Our results imply that the substorm current wedge is suppressed for $B_y<0$ in NH winter (rather than enhanced for $B_y>0$). 
While this effect is consistent with the observations of this paper, its physical mechanism still remains unknown.
Further studies are needed to better understand the physical mechanism behind the explicit $B_y$-effect in high-latitude geomagnetic activity.
The results of this paper are important for understanding and predicting space weather effects at high latitudes and for understanding the connection between long-term geomagnetic activity and solar wind variations.

%
%
%
%


%
%
%

%
%

%

\acknowledgments
We acknowledge the financial support by the Academy of Finland to the ReSoLVE Centre of Excellence (project no. 272157).
The solar wind data and the AL index were downloaded from the OMNI2 database (\texttt{http://omniweb.gsfc.nasa.gov/}).
The $K$-index of the Syowa station was downloaded from the National Institute of the Polar research, Japan at \texttt{http://polaris.nipr.ac.jp}.

%
%
%
%
%
%
%
%
%





\listofchanges

\end{document}